\documentclass[useAMS,usenatbib,usegraphicx]{mn2e}
\usepackage{ epstopdf, txfonts, color, soul,ulem, rotating}

\begin{document}

\title[Exoplanet radio emission variability]{Time-scales of close-in exoplanet radio emission variability}

\author[V. See et al.]
{V. See$^1$\thanks{E-mail: wcvs@st-andrews.ac.uk}, M. Jardine$^1$, R. Fares$^{2,1}$,  J.-F. Donati$^3$, C. Moutou$^{4,5}$\\
$^1$SUPA, School of Physics and Astronomy, University of St Andrews, North Haugh, KY16 9SS, St Andrews, UK\\
$^2$INAF-Osservatorio astrofisico di Catania, Via Santa Sofia 78, 95123 Catania, Italy\\
$^3$LATT–UMR 5572, CNRS and University P. Sabatier, 14 Av. E. Belin, F-31400 Toulouse, France\\
$^4$3 Canada-France-Hawaii Telescope Corporation, CNRS, 65-1238 Mamalahoa Hwy, Kamuela HI 96743, USA\\
$^5$4 Aix Marseille Université, CNRS, LAM (Laboratoire d'Astrophysique de Marseille) UMR 7326, 13388, Marseille, France}

\maketitle

\begin{abstract}
We investigate the variability of exoplanetary radio emission using stellar magnetic maps and 3D field extrapolation techniques. We use a sample of hot Jupiter hosting stars, focusing on the HD 179949, HD 189733 and $\tau$ Boo systems. Our results indicate two time-scales over which radio emission variability may occur at magnetised hot Jupiters. The first is the synodic period of the star-planet system. The origin of variability on this time-scale is the relative motion between the planet and the interplanetary plasma that is co-rotating with the host star. The second time-scale is the length of the magnetic cycle. Variability on this time-scale is caused by evolution of the stellar field. At these systems, the magnitude of planetary radio emission is anticorrelated with the angular separation between the subplanetary point and the nearest magnetic pole. For the special case of $\tau$ Boo b, whose orbital period is tidally locked to the rotation period of its host star, variability only occurs on the time-scale of the magnetic cycle. The lack of radio variability on the synodic period at $\tau$ Boo b is not predicted by previous radio emission models, which do not account for the co-rotation of the interplanetary plasma at small distances from the star. 

\end{abstract}


\section{Introduction}
Within the solar system, the magnetised planets are all known to be sources of intense non-thermal radio emission \citep{Zarka2007}. The emitted power is proportional to the power incident on the planet's magnetosphere. Surprisingly, the conversion efficiency between incident and emitted power is roughly constant for every planet. This relation is encapsulated by the radiometric Bode's law \citep{Desch1984,Farrell1999,Zarka2001,Zarka2007} with Jupiter having the strongest radio emission of the solar system planets.

In analogy with the solar system, it is thought that magnetised exoplanets may emit in the radio wavelengths. However, some stellar systems contain short-period Jupiter-mass planets \citep{Bayliss2011,Wright2012}. Due to the small star-planet distances ($<$0.1 AU), the amount of energy from the solar wind incident on so called `hot Jupiters' is higher and the magnitude of radio emission is expected to be correspondingly larger \citep{Stevens2005}.

Numerous authors have modelled exoplanetary radio emission with the aim of identifying the most promising targets for future observations \citep[e.g.][]{Lazio2004,Griessmeier2007b,Jardine2008}. Despite repeated attempts, there have been no confirmed exoplanetary radio emission detections to date \citep{Bastian2000,Ryabov2004,Lazio2007,Smith2009,Etangs2009,Etangs2011,Hallinan2013,Sirothia2014}. \citet{Bastian2000} offer several possible reasons for the lack of detection. The observations may have been made at a frequency different to that of the radio emission or could simply have lacked the sensitivity required. Alternatively, the radio emission may not have been beaming towards Earth. These explanations do not rule out the existence of radio emission, just our ability to detect it. The authors also argue that these systems could lack a source of keV electrons necessary to produce the emission. The lack of a detection would then be due to a genuine lack of emission. Lastly, the authors discuss the sporadic nature of radio emission. They argue that emissions above their detection threshold may exist but, due to emission variability, it was not detectable at the time of their observations.

With few exceptions \citep{Fares2010,Vidotto2012,Vidotto2015}, emission variability has been neglected in radio emission models to date. Most assume stellar magnetic fields and winds that are both steady and isotropic. However, observations of the solar wind \citep[e.g.][]{Gosling1996}, and modelling of stellar magnetic fields \citep[e.g.][]{Petit2008} and winds \citep[e.g.][]{Vidotto2009} show that this is not the case. {Indeed, the dynamic nature of the solar wind is one of the factors, among others, known to affect the intensity of radio activity at Jupiter and Saturn \citep{Zarka1998,Gurnett2004,Crary2005}.

In this paper we model the planetary radio emission from seven planet hosting systems, whose magnetic fields have been mapped by \citet{Fares2013}. In particular, we focus on HD 179949, HD 189733 and $\tau$ Boo. We incorporate these published magnetic maps into our model and consider the time-scales over which planetary radio emission variability might be expected. 

The rest of the paper will be structured as follows. In section \ref{sec:SystemModelling}, we outline our radio emission model. The sample of hot Jupiter hosting stars is discussed in section \ref{sec:SystemProperties}. In section \ref{sec:Results}, we discuss the results of the radio emission modelling for HD 179949b, HD 189733b and $\tau$ Boo b, and the rest of the sample and concluding remarks are given in section \ref{sec:Conclusion}.

\section{Radio emission model}
\label{sec:SystemModelling}
The model of exoplanetary radio emission employed in this work is adapted from the model presented by \citet{Jardine2008} and reproduces well the observed `radiometric Bode's law'. Radio emissions originate from a population of electrons that have been accelerated in the current sheet that forms where the stellar and planetary magnetic fields interact. The power of the accelerated electrons is given by

\begin{equation}
	P_{e}=V_{cs}\dot{n}_{run}K,
	\label{eq:POut}
\end{equation}
where $V_{cs}$ is the volume of the current sheet, $\dot{n}_{run}$ is the rate at which runaway electrons are generated, and $K$ is the characteristic energy to which each electron is accelerated\footnote{\citet{Jardine2008} state the power in the accelerated electrons is given by $\pi R_m^2 v N_{run} K$. However, the equation the authors give for the runaway electron number density, which they denote as $N_{run}$ (their Eq. 13), is actually the rate at which runaway electrons are generated per unit volume with units of m$^{-3}$s$^{-1}$. The correct form is given by our Eq. (\ref{eq:POut}). This modification does not affect the scalings, and hence conclusions, drawn by these authors.}. We take the acceleration region to be the size of the planetary magnetosphere, $r_{ms}$, with a fixed aspect ratio, $\alpha$. Its volume is therefore given by $V_{cs} = \pi r_{ms}^2 \cdot \alpha r_{ms} = \pi \alpha r_{ms}^3$. The magnetosphere size, determined by pressure balance between the wind ram pressure, stellar magnetic pressure and planetary magnetic pressure, is given by

\begin{equation}
	r_{ms} = \left(\frac{B^2_{pl}}{2\mu_0n_{wind}m_{p}v_{eff}^2+B^2_{\star}(r_{orb})}\right)^{1/6}r_{pl},
	\label{eq:rMag}
\end{equation}
where $B_{pl}$ is the planetary magnetic field strength, $\mu_0$ is the permeability of free space, $n_{wind}$ is the wind number density, $m_p$ is the proton mass, $v_{eff}$ is the effective wind speed, $B_{\star}(r_{orb})$ is the stellar field strength at the orbit of the planet and $r_{pl}$ is the planetary radius. The electric field, $E$, in the acceleration region determines the rate of runaway electron generation and has a dependence given by

\begin{equation}
	\frac{E}{E_D}\propto\frac{v_{eff}B_{\star}(r_{orb})}{n_{cs}},
	\label{eq:EField}
\end{equation}
where $E_D$ is the Dreicer field, $v_{eff}$ is the effective wind velocity, and $n_{cs}$ is the density within the acceleration region. \citet{Jardine2008} showed that the radiometric Bode's law for the Solar System planets can be reproduced if the density within the current sheet is enhanced by a factor, $f_{comp}$, when compared to the solar wind density due to compression within this region, such that $n_{cs} = f_{comp}n_{wind}$. We direct the interested reader to appendix \ref{app:Dreicer} for further details of the parameters $\dot{n}_{run}$, $E$, and $E_D$. Finally, the radio flux density received at Earth, $\Phi$, is given by 

\begin{figure*}
	\begin{center}
	\includegraphics[width=\textwidth]{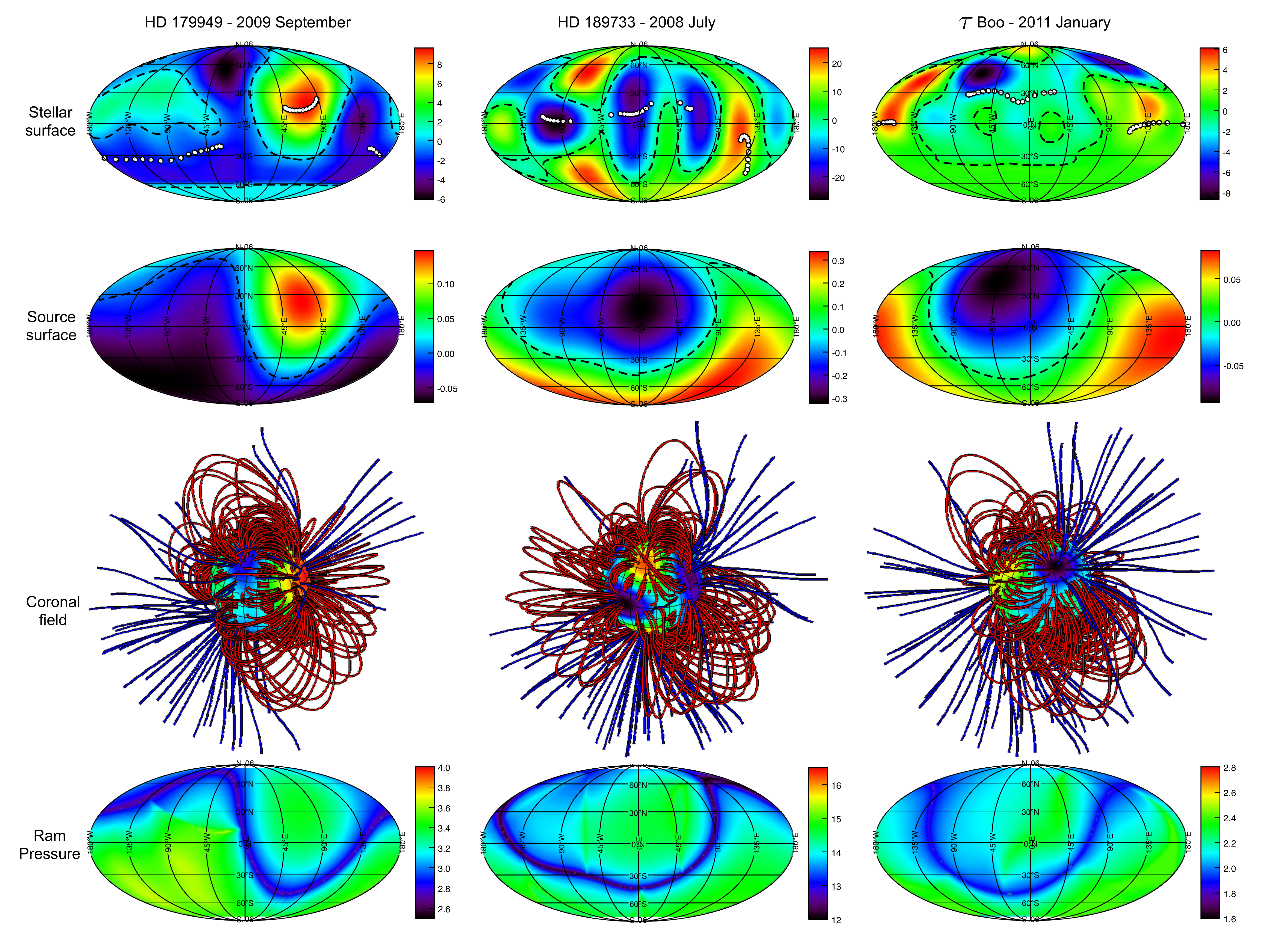}
	\end{center}
	\caption{Examples of the magnetic field geometries of HD 179949 (2009 September), HD 189733 (2008 July), and $\tau$ Boo (2011 January). For each star, maps of the radial magnetic field at the stellar surface (first row) and source surface (second row) are shown. The colour bars indicate the magnetic field strengths in Gauss. Polarity inversion lines are shown with dashed lines. White dots on the surface maps indicate the foot points of the field lines that connect with the planet around the orbit. The large-scale 3D coronal field structure between the stellar surface and the source surface is also shown (third row). Closed and open field lines are coloured in red and blue respectively. Finally, the stellar wind ram pressure, for all latitudes and longitudes, at the orbit of the planet are shown in units of $10^{-6}$ dyne cm$^{-2}$ (fourth row). All the stars display a complex field at the photosphere but a dipole dominated field exists at the source surface because higher order spherical harmonics decay more quickly with distance above the stellar surface. Three dimensional structure exists in the ram pressure and the stellar magnetic pressure (which has a structure similar to that seen in the second row).}
	\label{fig:FieldStructure}
\end{figure*}

\begin{equation}
	\Phi=\frac{P_{rad}}{\Omega d^2\delta f}
	\label{eq:FluxDensity}
\end{equation}
where $P_{rad}$ is the radio emission power, $\Omega$ is the solid angle into which the emission is beamed, $d$ is the distance to the system from Earth, and $\delta f$ is the bandwidth given by the maximum cyclotron frequency, $eB_{pl}/2\pi m_e$ \citep{Griessmeier2007a}. Here, $e$ and $m_e$ are the electron charge and mass respectively. Laboratory experiments have shown that roughly 10\% of the power in the accelerated electrons is converted to radio emission such that $P_{rad}=0.1P_{e}$. For this study, we adopt values of $K=$10KeV, $\alpha=10^{-8}$, $\Omega$=1.6sr and $f_{comp}$=15 which are characteristic values for these parameters, and Jovian parameters for the exoplanet. 

Two factors control the magnitude and variability of radio emissions at each system. The wind density determines how large a pool of electrons are available for acceleration within the current sheet while the stellar magnetic field determines the amount of energy available to accelerate those electrons. Though the wind density decreases as a function of orbital distance, roughly as $r^{-2}$ once the wind is terminal, it does not vary significantly around the planetary orbit within our model. Instead, it is the stellar magnetic field structure that induces variability in the radio emissions.

A feature of this model is that the radio flux density calculated by Eq. (\ref{eq:FluxDensity}) is independent of the assumed planetary field strength. A stronger field results in a larger planetary magnetosphere (Eq. (\ref{eq:rMag})). Since the volume of the current sheet is related to the magnetosphere size, $V_{cs} \propto r_{mp}^3$, a larger pool of electrons is available for acceleration (Eq. (\ref{eq:POut})). The increased power is offset by increased loses, associated with a higher bandwidth, as the radio signal propagates to Earth (Eq. (\ref{eq:FluxDensity})). Since the frequency of radio emission depends on the planetary magnetic field strength \citep{Zarka2007}, which are observationally unconstrained, our model does not make a prediction on the frequency of radio flux density at Earth, just its magnitude. 

The power of the accelerated electrons depends on the size of the magnetosphere, the effective velocity and the local electric field. These are determined by the large-scale flow and magnetic field. We discuss these aspects of the model in more detail in the following subsections.

\subsection{Stellar magnetic field extrapolation}
\label{subsec:FieldExtrapolation}
In this section, we describe the process by which the 3D magnetic field structure of cool stars can be determined using a potential-field source surface (PFSS) approach \citep{AltschulerNewkirk1969}. The magnetic field is assumed to be in a potential state ($\underline{\nabla} \times \underline{B} = 0$) which allows it to be defined in terms of a scalar potential, $\underline{B}=-\underline{\nabla} \psi$. Since no magnetic monopoles exist in nature, $\underline{\nabla} \cdot \underline{B} = 0$, Laplace's equation arises naturally, $\nabla^2 \psi = 0$. Solving Laplace's equation in terms of spherical harmonics gives us the coefficients of the associated Legendre polynomials, $P_{lm}$, from which the three magnetic field components at any point in the corona can be determined:

\begin{equation}
	B_r=-\sum\limits_{l=1}^N \sum\limits_{m=1}^l [la_{lm} r^{l-1} - \left(l+1\right) b_{lm} r^{-\left(l+2\right)}] P_{lm} \left(\cos \theta \right) e^{im\phi}
	\label{eq:Br}
\end{equation}

\begin{equation}
	B_{\theta}=-\sum\limits_{l=1}^N \sum\limits_{m=1}^l [a_{lm}r^{l-1} - b_{lm} r^{-\left(l+2\right)}]\frac{d}{d\theta} P_{lm} \left(\cos \theta \right) e^{im\phi}
	\label{eq:Btheta}
\end{equation}

\begin{equation}
	B_{\phi}=-\sum\limits_{l=1}^N \sum\limits_{m=1}^l [a_{lm}r^{l-1} - b_{lm} r^{-\left(l+2\right)}] P_{lm} \left(\cos \theta\right) \frac{im}{\sin \theta} e^{im\phi}
	\label{eq:Bphi}
\end{equation}
where $a_{lm}$ and $b_{lm}$ are the amplitudes of the spherical harmonics, $l$ is the spherical harmonic degree and $m$ is the order or `azimuthal number'. 

To determine the values of $a_{lm}$ and $b_{lm}$, we apply two boundary conditions. Magnetic maps reconstructed using Zeeman-Doppler imaging (ZDI) are used to set the first boundary condition at the stellar surface, $r_{\star}$. Examples of such surface maps are shown in Fig. \ref{fig:FieldStructure} (top row) for HD 179949 \citep{Fares2012}, HD 189733 \citep{Fares2010} and $\tau$ Boo \citep{Fares2012}. The second boundary condition is set at the source surface, $r_{ss}$, which represents the limit of coronal confinement. At the source surface, the magnetic field is forced to be purely radial, i.e. $B_{\theta}=B_{\phi}=0$ \citep{AltschulerNewkirk1969,Jardine2002}. Beyond the source surface, the field remains purely radial, decaying as an inverse square law. Physically, the source surface represents the location where the gas pressure is large enough to blow open closed coronal loops. We adopt a source surface distance of $r_{ss} = 3.4r_{\star}$. Since higher order spherical harmonic modes decay more rapidly with distance away from the stellar surface (Eqs. \ref{eq:Br} - \ref{eq:Bphi}), the field structure is predominantly dipolar at the source surface for these stars. This is true despite the more complex fields evident at the stellar surfaces shown in Fig. \ref{fig:FieldStructure}.

To carry out the field extrapolation, we use a modified version of the global diffusion code of \citet{vanB1998}. The PFSS approach has previously been used to study a variety of stars \citep{Jardine2002,Gregory2006,Lang2012,Johnstone2014}. Examples of the 3D extrapolation for HD 179949, HD 189733, and $\tau$ Boo are shown in Fig. \ref{fig:FieldStructure} as well as the field geometry at the source surface.

\subsection{Stellar wind}
\label{subsec:Wind}
Equation (\ref{eq:rMag}) requires us to estimate two properties of the stellar wind. These are the effective wind velocity, i.e. the speed at which the interplanetary plasma impinges on the planetary magnetosphere from the reference frame of the planet, and the density of the wind. Both of these quantities may vary around the orbit of the planet.

For short-period planets, the effective velocity, $v_{eff}$, has a component from the wind velocity, $v_{wind}$ and a component in the azimuthal direction, $v_{az}$, due to the orbital motion of the planet. The effective velocity is calculated by adding these components in quadrature,

\begin{equation}
	v_{eff}^2 = v_{az}^2 + v_{wind}^2. 
	\label{eq:vEff}
\end{equation}

At the small orbital distances where hot Jupiters exist, the interplanetary plasma will be corotating with the star. The azimuthal velocity component is therefore given by $v_{az} = 2\pi r_{orb}/P_{syn}$ where $r_{orb}$ is the orbital radius and $P_{syn}=\frac{P_{rot}P_{orb}}{P_{rot}-P_{orb}}$ is the synodic period of the planet with respect to the rotation of the star. $P_{rot}$ and $P_{orb}$ are the stellar rotation and planetary orbital periods respectively. In the rotating frame of the host star, it takes one synodic period for a planet to return to its starting position. Consequently,  radio emissions are periodic with a time-scale given by the synodic period \citep{Fares2010}.

We calculate $v_{wind}$ using a two step process. The wind speed is estimated at the source surface radius first. Solar wind speeds are known to correlate with the amount of field line divergence \citep{Levine1977,Wang1990}. \citet{Wang1991} showed that such a correlation is plausible provided that the amount of Alfven wave energy flux is constant within open flux tubes. Building on the work of \citet{Wang1990}, \citet{Arge2000} establish a continuous empirical relation between the magnetic expansion factor, $f_s$, and the solar wind velocity at the source surface of the Sun,

\begin{equation}
	v_{wind}\left(r_{ss}\right)=267.5+\frac{410}{f_s^{2/5}} [\mathrm{ms^{-1}}].
	\label{eq:WSAVelocity}
\end{equation}
The magnetic expansion factor is given by

\begin{equation}
	f_s=\left(\frac{r_{\odot}}{r_{ss}}\right)^2 \frac{B(r_{\odot})}{B(r_{ss})},
	\label{eq:ExpansionFactor}
\end{equation}
where $B(r_{ss})$ is the magnetic field strength at a location on the source surface and $B(r_{\odot})$ is the magnetic field strength at the solar surface along the same magnetic field line. Since the so called Wang-Sheeley-Arge model is calibrated for the solar wind, we modify Eq. (\ref{eq:ExpansionFactor}), replacing $r_{\odot}$ with $r_{\star}$. $B(r_{\star})$ and $B(r_{ss})$ are determined in the coronal field extrapolation by eqs. (\ref{eq:Br})  - (\ref{eq:Bphi}).

The second step is to propagate $v_{wind}(r_{ss})$ out to $r_{orb}$. Above the source surface, we assume that the wind evolves according to the model of \citet{Parker1958}. The wind velocity is found by integrating the momentum equation for the wind,

\begin{equation}
	\rho m v_{wind}\frac{\partial v_{wind}}{\partial r}=-\frac{\partial}{\partial r}(\rho k_BT)-\rho m \frac{GM_{\star}}{r^2},
	\label{eq:ParkerWind}
\end{equation}
where $\rho$ is the mass density, $m$ is the average molecular mass, $T$ is the isotropic wind temperature and the other symbols have their usual meanings. The wind temperature is chosen to match the velocities at the source surface. The wind velocity at the orbit of the planet can then be determined using this temperature. Our approach to calculating the wind velocity allows the full 3D structure of the stellar magnetic field to be imprinted on the wind velocity. Thus the wind ram pressure varies around the planetary orbit in a way that is determined by the observed magnetic field geometry.

Presently, no direct methods of observing the very low density winds of low-mass stars exist. \citet{Jardine2008} use a scaled solar wind density in their model. These authors set the wind density to be $N_W=1.7 \times 10^{-20} kgm^{-3}$ at a distance of $r=215r_{\star}$ from the host star. The density at the exoplanet is then calculated according to mass conservation and assuming a wind that is terminal and radial, i.e. the wind density falls as $r^{-2}$. However, since we use a radially varying wind in this work, the density falls as $v_{wind}^{-1}r^{-2}$ in order to conserve mass. Additionally, we scale the wind density by a factor, $f_{mag}$, where $f_{mag}$ is the average foot point strength of the field lines that intersect the exoplanet's orbit normalised to the average Solar surface field (1G). Physically, $f_{mag}$ accounts for the denser winds of more active stars \citep{Mestel1987}. The density at the orbit of the planet is therefore given by

\begin{equation}
	n_{wind}(r) = 10^7f_{mag}\frac{v_{wind}(r=215r_{\star})}{v_{wind}(r)}\left(\frac{r_{\star}}{r_{\odot}}\right)^2\left(\frac{r}{1\textrm{AU}}\right)^{-2}.
	\label{eq:Density}
\end{equation}
In the last row of Fig. \ref{fig:FieldStructure}, we show the ram pressure of the wind on a shell at the orbital radius of the planet. The influence of the magnetic field structure can clearly be seen in the ram pressure structure.

\begin{figure*}
	\begin{center}
	\includegraphics[width=0.85\textwidth]{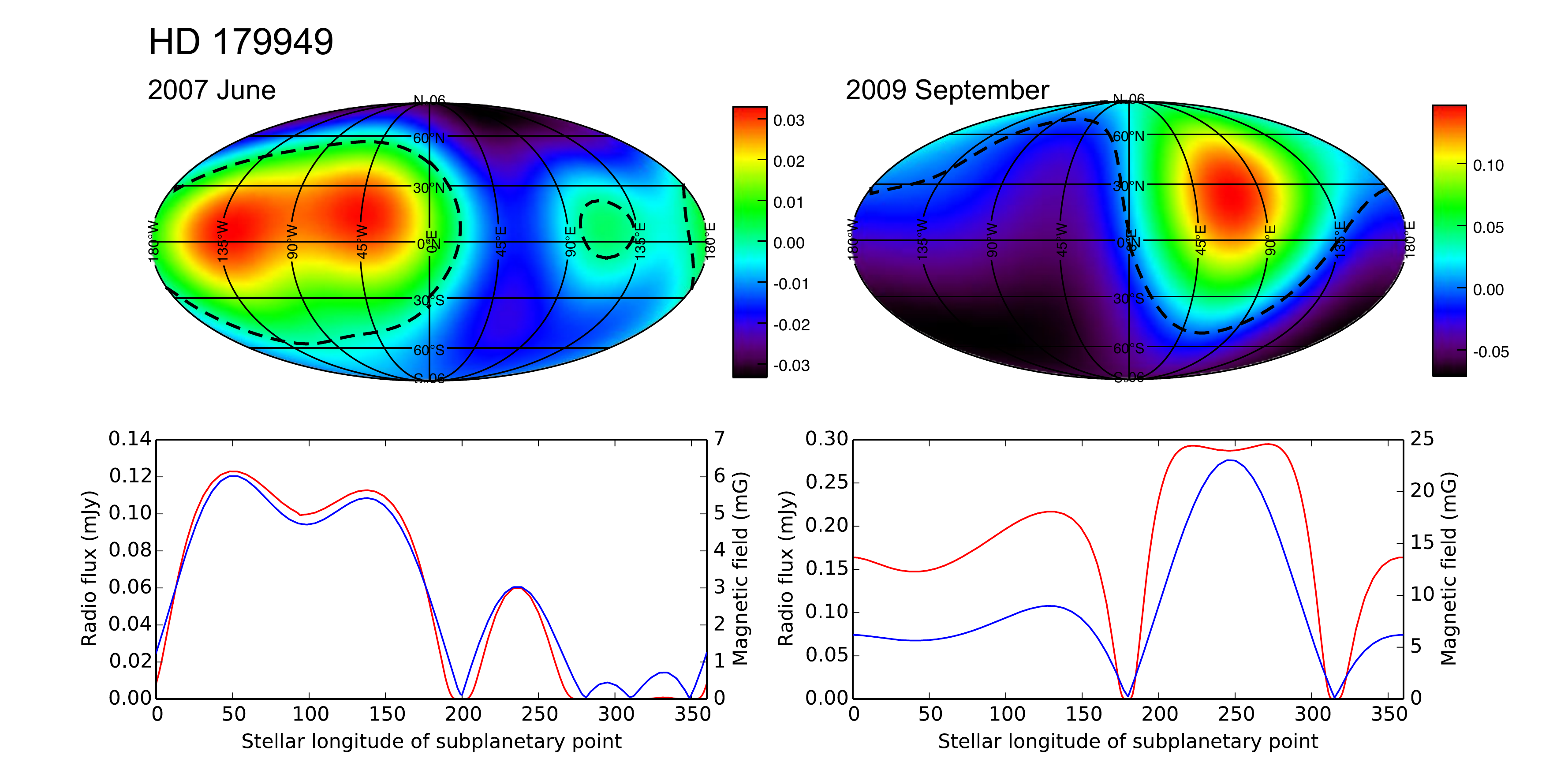}
	\end{center}
	\caption{Properties of the HD 179949 system during the 2007 June (left) and 2009 September (right) epochs. Top: Magnetic field maps of the source surface displayed in the same format as Fig. \ref{fig:FieldStructure}. Bottom: Predicted planetary radio flux (red) and stellar magnetic field strength along the planetary orbit (blue) as a function of the stellar longitude of the subplanetary point. Large drops in radio emission are evident as the planet crosses the polarity inversion line.}
	\label{fig:179949Flux}
\end{figure*}

\begin{table}
\begin{minipage}{80mm}
	\caption{Properties of the HD 189733, HD 179949 and $\tau$ Boo systems. Listed are the spectral type, stellar mass and radius, planetary orbital radius, stellar rotation period, planetary orbital and synodic periods and distance from the Sun. All values referenced from \citet{Fares2013} with the exception of $P_{syn}$ (calculated) and $d$ (referenced below).}
	\label{tab:StarProperties}
	\begin{tabular}{lccc}
		\hline
		Parameter  & HD 189733 & HD 179949 & $\tau$ Boo\\
		\hline
		SpT & K2V & F8V & F7V\\
		$M_{\star}$ $[M_{\odot}]$ & 0.82 & 1.21 & 1.34\\
		$r_{\star}$ $[r_{\odot}]$ & 0.76 & 1.19 & 1.42\\
		$r_{orb}$ [AU] & 0.031 & 0.0439 & 0.048\\
		$P_{rot}$ [days] & 12.5 & 7.6 & 3.31$^a$\\
		$P_{orb}$ [days] & 2.22 & 3.09 & 3.31\\
		$P_{syn}$ [days] & 2.7 & 5.2 & $\infty$\\
		$d$ [pc] & 19.3$^b$ & 27.1$^c$ & 15.6$^d$\\
		\hline
	\end{tabular}
    	\small
    	\begin{itemize}
    	\item[] $^a$\citet{Fares2013} use the equatorial rotation period of 3 days for $\tau$ Boo. We choose to use a rotation period 3.31 days, which is still consistent with the differential rotation exhibited by $\tau$ Boo, to explore the effects of a tidally locked planet.
    	\item[] $^b$\citet{Fares2010}, $^c$\citet{Butler2006}, $^d$\citet{Griessmeier2007a}
    	\end{itemize}
\end{minipage}
\end{table}

\section{Stellar system properties}
\label{sec:SystemProperties}
In this paper, we use the sample of seven hot Jupiter hosting stars presented by \citet{Fares2013} for which a ZDI map exists. In particular we focus on HD 179949, HD 189733 and $\tau$ Boo. Properties for these three systems are shown in Table \ref{tab:StarProperties} with details for the full sample available in \citet{Fares2013}. 

We choose these three systems for further discussion because they are often cited as promising targets for detectable exoplanetary radio emission \citep{Lazio2004,Griessmeier2007b,Jardine2008} and have been extensively studied in the literature. In addition, these stars allow a comparison of two critical factors that affect exoplanetary radio emission. The first is the spectral type of the host star which affects the nature of the stellar wind. HD 179949, an F8 dwarf, and HD 189733, a K2 dwarf, allow us to make this comparison. The second is the relative angular velocity of the planet's orbital motion and the stellar rotational motion which determines whether the planet moves relative to the stellar magnetic field. We can evaluate the effect of this factor with HD 179949b and $\tau$ Boo b. Both planets have host stars with similar spectral types but the latter system is known to be tidally locked whilst the former is not.

HD 189733b and $\tau$ Boo b are known to orbit in, or close to, the stellar equatorial plane. Making use of the Rossiter-McLaughlin effect \citep{Rossiter1924,Mclaughlin1924}, \citet{Triaud2009} determine the spin-orbit misalignment angle for HD 189733b to be $0.85^{\circ+0.32}_{-0.28}$. For $\tau$ Boo b, \citet{Brogi2012} find that the normal to the orbital plane is not significantly misaligned with the stellar rotation axis. For the purposes of this study, we will assume that both planets lie exactly in the stellar equatorial plane. For HD 179949b, \citet{Brogi2014} determine that the orbital inclination is $67.7^{\circ} \pm 4.3^{\circ}$. \citet{Fares2012} determine that the stellar inclination is roughly $60^{\circ}$ using ZDI. Inclinations determined from ZDI can have errors of 10$^{\circ}$ or greater \citep[e.g.][]{Petit2008}. We will assume a spin-orbit misalignment of 0$^{\circ}$ for HD 179949 since this is consistent with the stellar and orbital inclinations when considering the errors. \citet{Winn2010} note that hot Jupiter systems with hotter host stars ($T_{eff} > 6250$K) are more likely to have a large spin-orbit misalignment. Within our sample, HD 179949 and $\tau$ Boo are the hottest stars, with effective temperatures of 6168K and 6387K respectively \citep{Fares2013}, and are the most likely to have large spin-orbit misalignments according to the trend noted by \citet{Winn2010}. Nevertheless, we will proceed by assuming the planetary orbits are equatorial for the reasons given above. For the remaining five systems, the spin-orbit misalignments are unknown since the planets are not transiting. For the purposes of this study, we assume that they also orbit within the stellar equatorial plane. 

\begin{figure*}
	\begin{center}
	\includegraphics[width=0.85\textwidth]{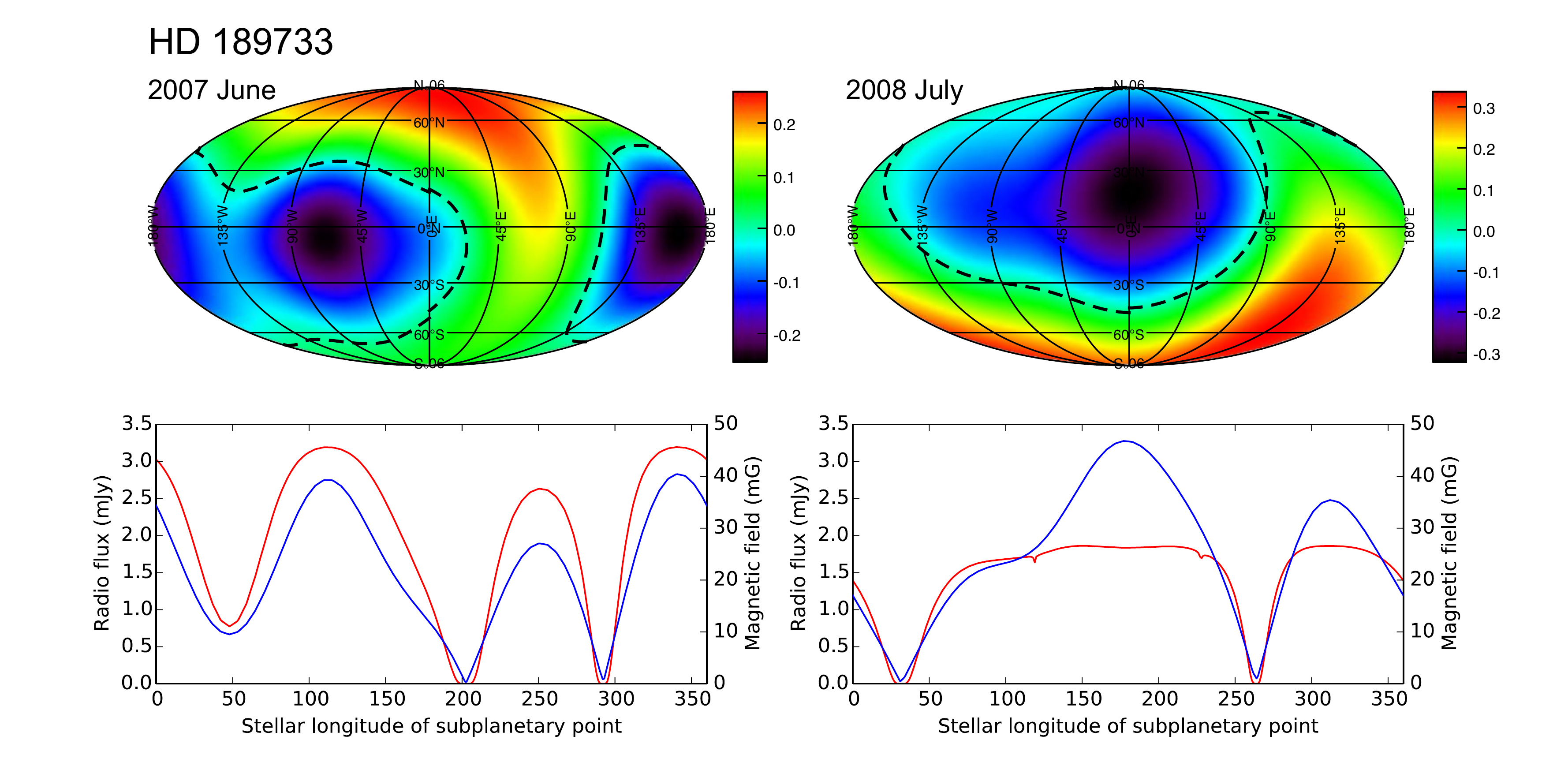}
	\end{center}
	\caption{Properties of the HD 189733 system during the 2007 June (left) and 2008 July (right) epochs. Data displayed in the same format as Fig. \ref{fig:179949Flux}.}
	\label{fig:189733Flux}
\end{figure*}

In terms of the geometry of the system, it is also important to consider the fact that the radio emission is anisotropic. The emissions are beamed into a cone which is aligned with the local planetary magnetic field \citep{Zarka2007}. If the magnetic dipole axis and planetary rotation axis are aligned, the radio emissions should be beamed in the direction of Earth at all times, if the geometry is favourable, or not at all. However, if there is a misalignment between the planetary magnetic dipole and rotation axes, then emissions, as observed from Earth, may be modulated on the planetary rotation period as the emission cone swings in and out of view. Given that we have no information about the planetary field alignment, we will assume the most favourable geometry, in which the radio emissions are always beamed towards Earth.

For each star in our sample, magnetic maps are reconstructed from spectropolarimetric observations. The observations were made using one of two high-resolution spectropolarimeters, either ESPaDOnS at the 3.6-m Canada-France-Hawaii Telescope (CFHT) on Mauna Kea or NARVAL at the 2-m T\'elescope Bernard Lyot (TBL) in Pic du Midi (France). Two magnetic maps are available each for HD 179949 \citep{Fares2012} and HD 189733 \citep{Fares2010}. For $\tau$ Boo, we use seven ZDI maps spanning four years \citep{Donati2008,Fares2009,Fares2013}. ZDI maps for the remaining four stars are available in \citet{Fares2013}. In Fig. \ref{fig:FieldStructure} we show examples of the magnetic field structure for HD 179949, HD 189733, and $\tau$ Boo. In each case, the radial component of the magnetic field at the stellar surface \& source surface is shown, as well as the 3D coronal field structure between these two boundaries. Though complex structure is evident at the surface of each of the stars, it is predominantly the dipole component that survives at the source surface. 

\begin{figure*}
	\begin{center}
	\includegraphics[width=\textwidth]{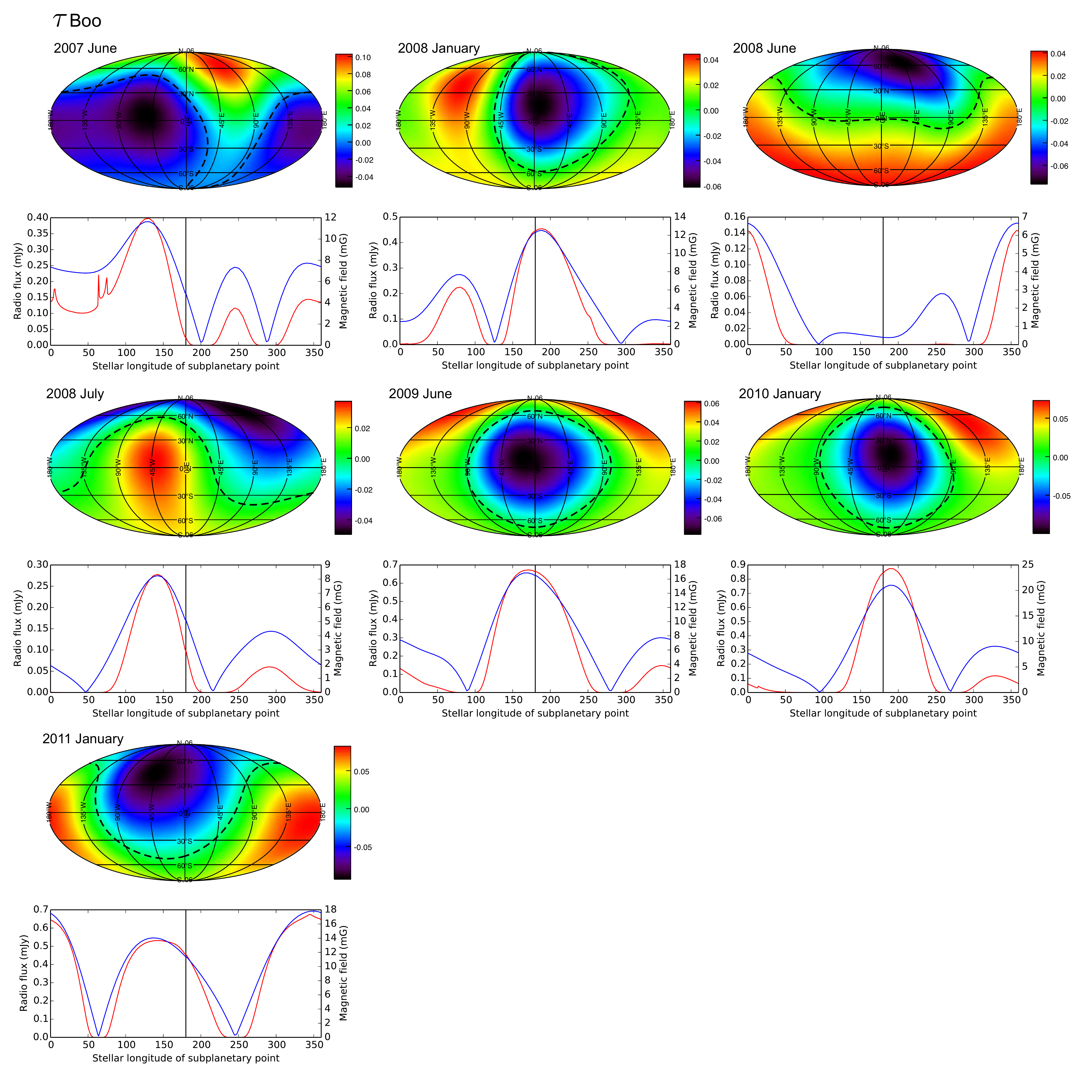}
	\end{center}
	\caption{Properties of the $\tau$ Boo system over seven epochs. Data is displayed in the same format as Figure \ref{fig:189733Flux}. The predicted radio flux displayed is that which could be expected if $\tau$ Boo b's subplanetary point sampled every stellar latitude. However it remains fixed at a single stellar longitude, 180$^{\circ}$, due to tidal locking. Black vertical lines are included to indicate this longitude. The expected radio flux, at any given epoch, is given by the intersection of the red and black lines. Figure \ref{fig:FixedTauBoo} shows how this flux varies with time.}
	\label{fig:TauBooFlux}
\end{figure*}

\section{Results}
\label{sec:Results}
\subsection{HD 179949}
\label{subsec:HD179949}
Figure \ref{fig:179949Flux} shows magnetic maps at the source surface of HD 179949 (top row). These maps are obtained using the extrapolation process described in section \ref{subsec:FieldExtrapolation} using magnetic maps of the stellar surface obtained during 2007 June and 2009 September \citep{Fares2012}. The colours represent the magnetic field strength in Gauss and the dashed lines represent polarity inversion lines, i.e. locations where $B_r=0$. To calculate the expected planetary radio emission, we require the local stellar magnetic field strength in the vicinity of HD 179949b as it orbits over one synodic period. This is obtained by considering the stellar magnetic field strengths in the equatorial plane of the source surface maps and calculating the magnetic field strength at the distance of the planet according to an inverse square law decay. The local field strength has been plotted in the bottom row of Fig. \ref{fig:179949Flux} in blue as a function of the stellar longitude of the planet. The corresponding predicted radio flux received at Earth, over one synodic period, is also plotted in red. Figure \ref{fig:179949Flux} has been formatted such that the stellar longitudes of the top and bottom rows are aligned.

Over each synodic period (5.2 days), HD 179949b exhibits highly variable radio emission that correlates strongly with the local stellar field strength. The variability can be mostly attributed to the orbital motion through the stellar magnetic field. Comparing the source surface maps with the predicted radio flux curves, negligible emission occurs as the planet crosses the polarity inversion lines. Conversely, the highest emission occurs when the planet is over the magnetic poles, i.e. when the field strength is strongest. This is true during both epochs, between which, the stellar field has evolved. 

As well as planetary radio emission, several authors have suggested that magnetic interactions may also induce enhanced chromospheric activity near the stellar surface \citep{Shkolnik2008,Lanza2009}. \citet{Fares2012} searched for this type of interaction at HD 179949 finding hints of activity enhancement modulated at the synodic period of the system. Given that enhanced chromospheric activity and planetary radio emission are both thought to occur as a result of magnetic interactions between a planet and host star, it is unsurprising that both would be modulated on the synodic period.

\subsection{HD 189733}
\label{sec:HD189733}
In Fig. \ref{fig:189733Flux}, we plot the same information as Fig. \ref{fig:179949Flux} but for the HD 189733 system. The ZDI maps used for the field extrapolations were presented by \citet{Fares2010}. Comparing Figs. \ref{fig:179949Flux} and \ref{fig:189733Flux}, we see that HD 179949b and HD 189773b both display strong variability, modulated on the synodic period of their respective systems during a given epoch. Given that both planets orbit stars where the dipole component of the magnetic field is tilted, the similarity is to be expected. The main difference between these systems is the magnitude of the radio flux density, smaller by roughly an order of magnitude at HD 179949b. This can be attributed to the lower photospheric magnetic field strength, the larger star-planet distance, and the greater distance from Earth for HD 179949.

Using the model presented by \citet{Griessmeier2007b}, \citet{Fares2010} have also predicted the radio emission expected from HD 189733b. The shape of their radio flux curve as a function of subplanetary point is in agreement with ours though the magnitude of their radio emission is higher, reaching the hundreds of mJy, due to a higher conversion efficiency from the Poynting flux to radio emission.

A number of authors have searched for radio emission from HD 189733b. While none produced positive detections, 3$\sigma$ upper limits of 2.1 mJy, 2mJy, 81mJy, and 160$\mu$Jy were found at frequencies of 150MHz \citep{Etangs2011}, 244MHz \citep{Etangs2009}, 307-347MHz \citep{Smith2009}, and 614MHz \citep{Etangs2009} respectively. Recalling that our model does not predict emission of a specific frequency, we can compare these limits to our predicted radio flux densities. Assuming our predictions are correct, it is unsurprising that \citet{Smith2009} were not able to detect any radio emissions given their upper limit of 81mJy. However, at the other three frequency bands, the upper limits are either comparable, or significantly lower, than our predictions. \cite{Etangs2009} discuss several reasons for the lack of detection and conclude that the most likely is because the emission was at a lower frequency.

\subsection{$\tau$ Boo}
\label{sec:TauBoo}
Similarly to Figs. \ref{fig:179949Flux} and \ref{fig:189733Flux}, we plot magnetic maps of the source surface and the predicted planetary radio emission for the $\tau$ Boo system in Fig. \ref{fig:TauBooFlux}. We use seven ZDI maps, with epochs spread over three years, for the field extrapolations of $\tau$ Boo \citep{Donati2008,Fares2009,Fares2013}. 

Unlike the previous systems, the orbital motion of $\tau$ Boo b is tidally locked to its host star \citep{Donati2008}. Consequently, the subplanetary point is fixed at one stellar longitude (180$^{\circ}$ in Fig. \ref{fig:TauBooFlux}). Since there is no relative motion between planet and host star, radio emission variability from planetary motions through the stellar field no longer exists. The only source of variability, in this model, is from dynamo driven evolution of the stellar field between epochs. It is worth highlighting that this behaviour would not be seen from models where the stellar rotational velocity is neglected, such that the $v_{az}$ component of $v_{eff}$ is set to the Keplerian velocity of the planet. If $v_{az}$ is non-zero, there must be some relative motion between the planet and the surround interplanetary plasma and magnetic field.

\begin{figure}
	\begin{center}
	\includegraphics[width=\columnwidth]{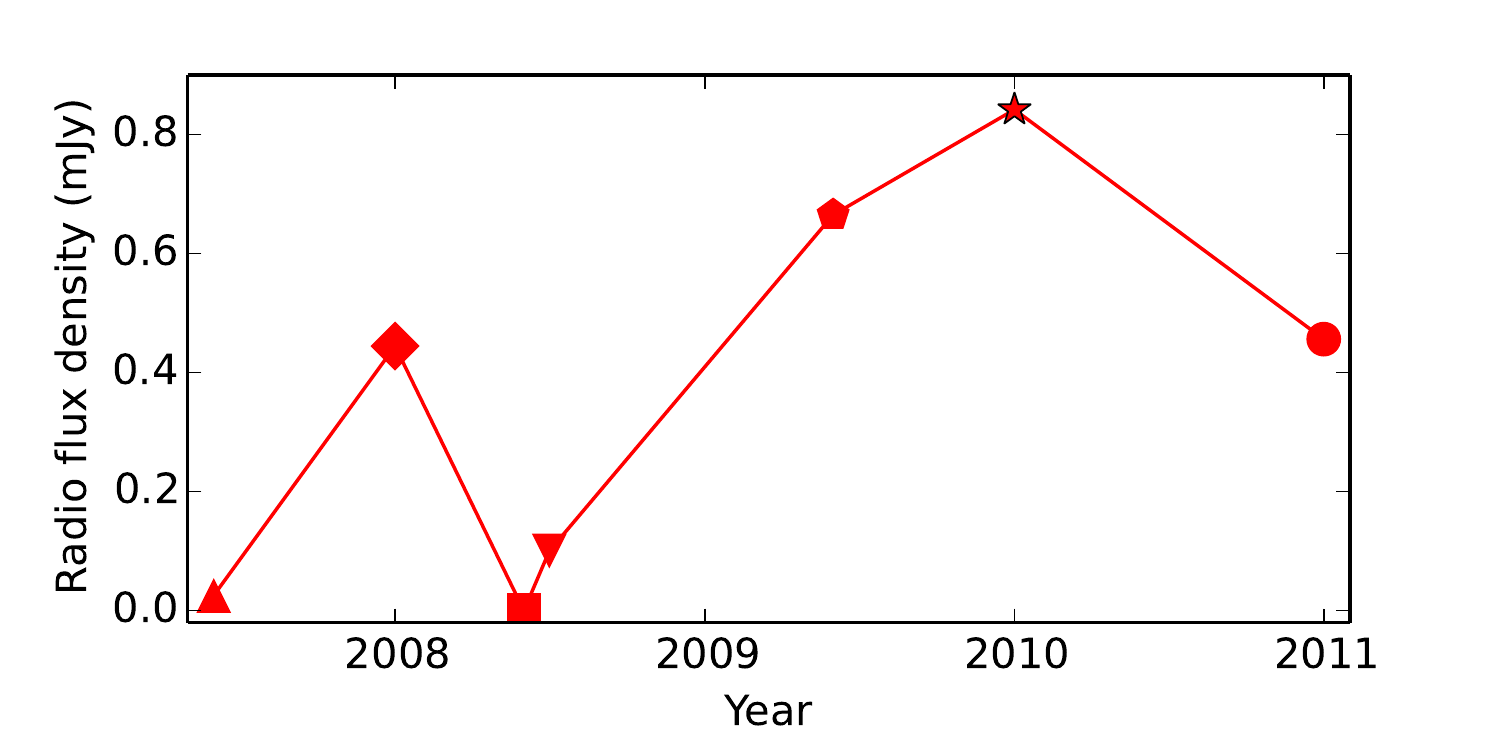}
	\end{center}
	\caption{Predicted radio flux of $\tau$ Boo b as a function of time. The connecting lines are visual aids only and should not be interpreted as a form of interpolation. Each epoch is plotted with a different symbol to allow easier comparison with Figs. \ref{fig:PoleDrift} and \ref{fig:AngSep}. In contrast to HD 189733, where the principle source of variability can be attributed to the orbital motion of the planet, variations in radio flux are caused by dynamo drive1n evolution of the stellar magnetic field.}
	\label{fig:FixedTauBoo}
\end{figure}

\begin{figure}
	\begin{center}
	\includegraphics[width=\columnwidth]{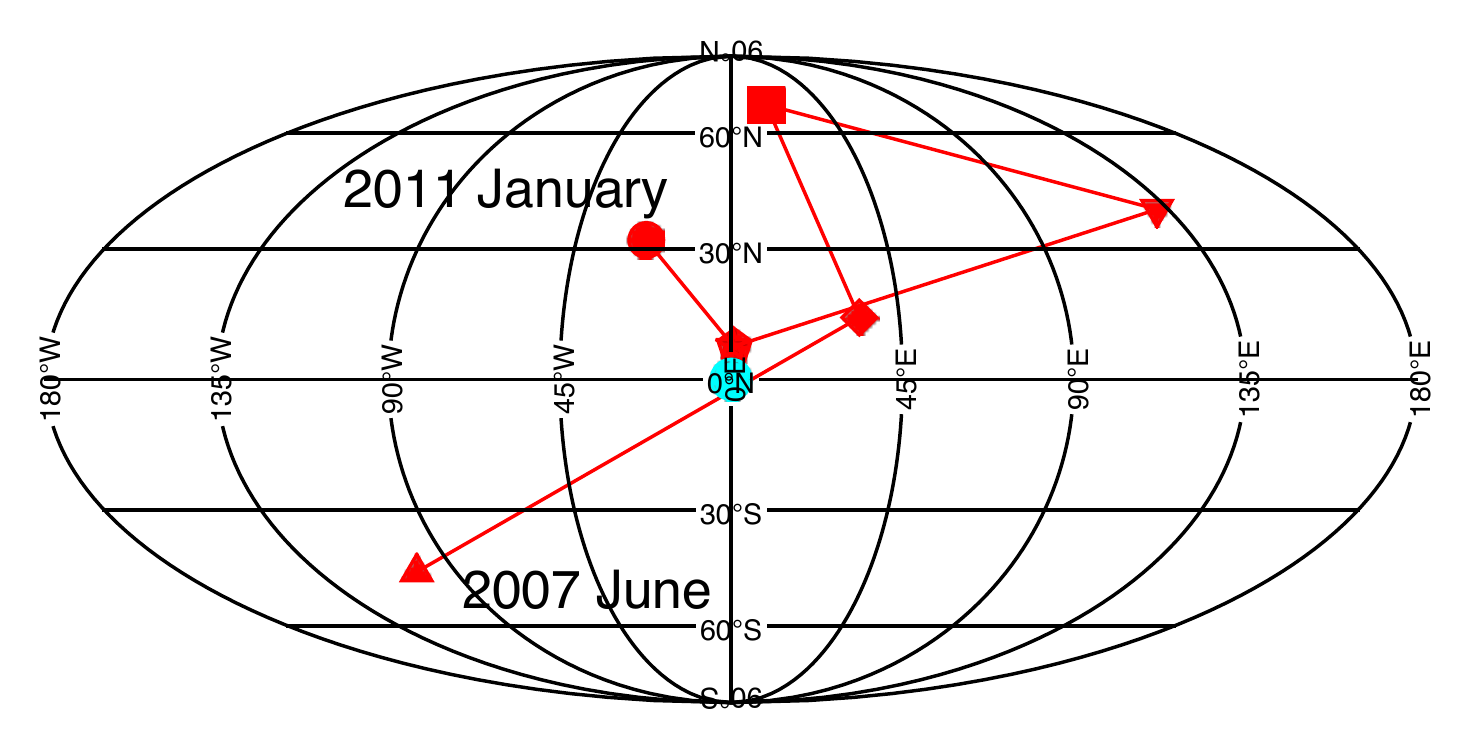}
	\end{center}
	\caption{Location of the negative pole of the dipole component of $\tau$ Boo's magnetic field during each epoch. Lines connecting contiguous epochs are visual aids only and should not be interpreted as a form of interpolation. For each epoch, the symbol used is the same as that used in Fig. \ref{fig:FixedTauBoo}. The negative poles during 2009 June and 2010 January epochs are very close (latitude $\sim$8$^{\circ}$ and longitude $\sim$180$^{\circ}$) and only appear to be a single data point in this plot.}
	\label{fig:PoleDrift}
\end{figure}

\begin{figure}
	\begin{center}
	\includegraphics[width=\columnwidth]{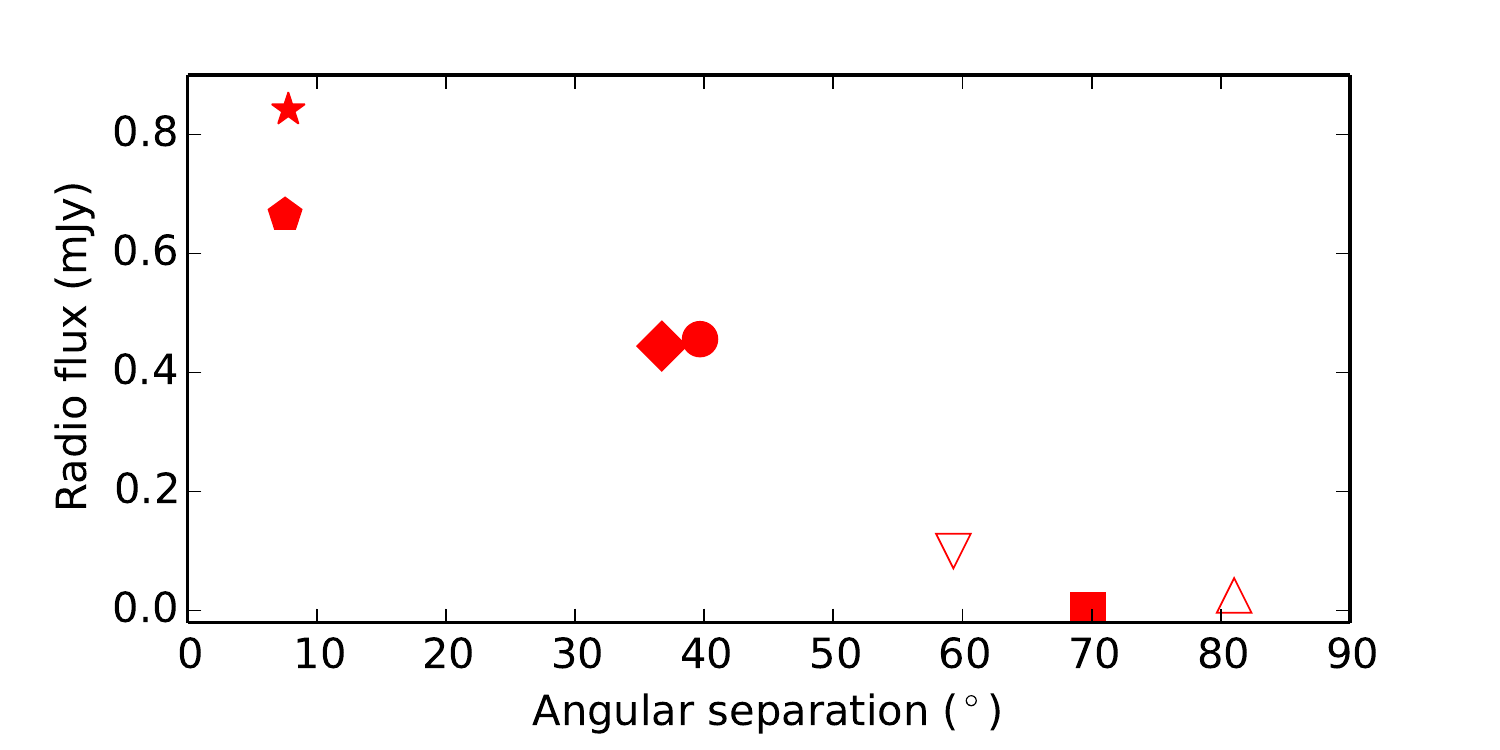}
	\end{center}
	\caption{Radio flux as a function of angular separation between the subplanetary point and the closest magnetic pole of $\tau$ Boo during each epoch. Symbol shapes correspond to those used in Figs. \ref{fig:FixedTauBoo} and \ref{fig:PoleDrift}. Filled symbols are used if the nearest pole is negative (the points plotted in Fig. \ref{fig:PoleDrift}) and unfilled symbols are used if the nearest pole is positve (antipodal to those in Fig. \ref{fig:PoleDrift}). A trend of decreasing radio flux can clearly be seen for increasing angular separation from the nearest stellar magnetic pole.}
	\label{fig:AngSep}
\end{figure}

Given the lack of relative motion between host star and planet, we would not expect to see the entirety of the predicted radio curves in Fig. \ref{fig:TauBooFlux}. Rather, these curves represent the expected radio flux if $\tau$ Boo b were not locked to its host star. In reality, we would only expect to see the radio flux indicated at 180$^{\circ}$ during each epoch, which has been highlighted by black vertical lines.  In Fig. \ref{fig:FixedTauBoo}, we show the predicted radio flux from $\tau$ Boo b as a function of time by collecting into one plot the radio fluxes, at 180$^{\circ}$ for each epoch, from Fig. \ref{fig:TauBooFlux}. A different symbol is used for each epoch to allow easier comparison between Figs. \ref{fig:FixedTauBoo} - \ref{fig:AngSep}. This plot demonstrates that variability occurs over months/years, i.e. the time-scale of the field evolution. The predicted radio fluxes are of the same order of magnitude as those predicted by \citet{Vidotto2012} despite the differing models and assumptions used.

The radio flux variations in Fig. \ref{fig:FixedTauBoo} occur due to evolution of the stellar field. The magnitude of the radio flux is highest when $\tau$ Boo b is over a magnetic pole, e.g. 2008 January, and weakest when it is over a magnetic cusp, e.g. 2008 June. Figure \ref{fig:PoleDrift} shows the location of the negative pole of the dipole component ($l$=1 mode) during each epoch (red markers) and the subplanetary point (blue marker). Figure \ref{fig:AngSep} shows the predicted radio flux as a function of angular separation between $\tau$ Boo b's subplanetary point and the nearest magnetic pole during each epoch (filled symbols if the nearest pole is negative and unfilled if it is positive). A clear trend of decreasing radio flux for increasing angular separation to the nearest magnetic pole is apparent. This result highlights the importance of knowing the three dimensional structure of the stellar magnetic field and not just the structure in the orbital plane of the exoplanet. It also argues for further ZDI observations of $\tau$ Boo which may allow us to determine any periodic behaviour in its large-scale magnetic field. If the evolution of the dipole component could be predicted, some estimate of the radio flux in the future could be made. 

\subsection{Rest of sample}
\label{subsec:RestOfSam}
From the sample of \citet{Fares2013}, four further stars had detectable magnetic fields with one ZDI map existing for each. None of these planets are tidally locked to their host star. If any of these planets are radio emitters, all should show variability as a result of their orbital motions. Table \ref{tab:RestOfSam} shows the predicted maximum planetary radio flux received at Earth per synodic period and the synodic period for each of these systems. None of these planets are particularly strong radio emitters. \citet{Vidotto2015} have also calculated the radio fluxes emitted from these exoplanets. With the exception of HD 130322, they find fluxes which are roughly comparable (within a factor of 1-4). The reasons we find such low radio fluxes are the same as those given to explain HD 179949b's lower radio flux density compared to HD 189733b's, namely low photospheric magnetic field strengths, large star-planet distances, and large distances to Earth. The combination of these factors means that these systems are not ideal targets for future planetary radio observations. 

\begin{table}
\begin{minipage}{80mm}
	\caption{Maximum predicted planetary radio fluxes for the rest of the systems presented by \citet{Fares2013}.}
	\label{tab:RestOfSam}
	\begin{tabular}{lcccc}
		\hline
		HD No. & 102195 & 130322 & 46375 & 73256\\
		\hline
		$\Phi_{max}$ [mJy] & 0.062 & $6.0 \times 10^{-4}$ & 0.018 & 0.19\\
		$P_{syn}$ [days] & 6.2 & 18.2 & 3.3 & 3.1\\
		\hline
	\end{tabular}
\end{minipage}
\end{table}

\section{Discussion \& Conclusion}
\label{sec:Conclusion}

We have presented a planetary radio emission model that incorporates a realistic geometry for the large-scale stellar magnetic field. Additionally, our model estimates the effective wind velocity, $v_{eff}$, more accurately than existing models in the literature. An accurate estimate should include a wind component, $v_{wind}$, that accelerates radially, and an azimuthal component, $v_{az}$, that accounts for the exoplanetary orbital motion relative to the stellar rotation. \citet{Stevens2005} assumes $v_{eff}=v_{wind}=400$kms$^{-1}$ but do not include an azimuthal component. This is likely to be an overestimate of $v_{wind}$ and an underestimate of $v_{az}$. \citet{Jardine2008} assume $v_{eff}=v_{az}$ for planets orbiting close to their host star. These authors account for the stellar rotation but do not include a wind component. \citet{Griessmeier2007a} incorporates a radially accelerating wind component and an azimuthal component. However, they do not account for the stellar rotation and so overestimate values for $v_{az}$ and hence $v_{eff}$. These examples demonstrate the need for careful consideration of the stellar system dynamics when calculating the effective velocity parameter. 

Focusing on the HD 179949, HD 189733 and $\tau$ Boo planetary systems, we looked at the time-scales on which exoplanetary radio emissions may vary. We find that planetary radio emissions are strongly dependent on local stellar magnetic field strength along the exoplanetary orbit. Therefore, exoplanetary radio emissions could be used as a probe of the stellar magnetic field geometry. 

In general, the height of magnetic loops on the surface of cool stars are comparable to their foot point separation. Therefore, higher order multipoles will decay more rapidly with height above the stellar surface. These higher order modes will contribute to the variability in the radio emissions of exoplanets that are orbiting close to the stellar surface of their host star. However, exoplanets at larger orbital radii, such as those in our sample, will predominantly experience the dipole component of the stellar field. Emission maxima exist when the planet is above a pole while a minima would be expected inside a magnetic cusp. In general, the magnitude of radio emission is anticorrelated with the angular separation between the subplanetary point and the closest magnetic pole.

We find two time-scales for variability. The shorter time-scale is the synodic period of the planetary system and is caused by the orbital motion of the planet. From the reference frame of the planet, the local stellar magnetic field strength is constantly varying causing corresponding variations in the radio emission, typically on the order of days. This type of behaviour, which is exhibited by the HD 179949 and HD 189733 systems, has implications for future observations. It is similar to the behaviour described by \citet{Llama2013} with respect to observing early ingresses for transiting exoplanets. In both cases, unfavorable wind conditions may lead to a lack of detection and highlights the importance of considering the field/wind conditions when planning future observations. 

The longer time-scale is most apparent in systems where the orbital and stellar rotation periods are similar in length. This results in a very long synodic period. In the most extreme case, when the stellar rotation period and orbital period are equal, there is no relative motion between planet and the stellar atmosphere in the azimuthal direction. For such systems, e.g. $\tau$ Boo, variability only occurs as a result of stellar magnetic field evolution. Previous models which estimate the azimuthal component of the effective velocity to be the Keplerian speed of the planet do not predict this type of behaviour for tidally locked systems. Indeed, by definition, the effective velocity in such systems cannot have a non-zero azimuthal component. One would expect variability at locked systems to occur over a characteristic time given by the magnetic cycle period. Non tidally locked systems will display variability on this time-scale as well but the time-scale for variability as a result of orbital motions will be much shorter.

Future exoplanetary radio emission detection attempts should focus on exoplanets orbiting close to stars with with strong magnetic fields and which are close to Earth. Additionally, the exoplanet would ideally have a large orbital velocity in the rotating frame of its host star. This would result in a higher azimuthal component of the effective wind velocity and hence higher Poynting flux. 

\section*{Acknowledgements}
The authors are thankful for helpful comments from an anonymous referee. V.S. acknowledges the support of an STFC studentship. R.F. acknowledges support from STFC consolidated grant ST/J001651/1. This work is based on observations obtained with ESPaDOnS at the CFHT and with NARVAL at the TBL. CFHT/ESPaDOnS are operated by the National Research Council of Canada, the Institut National des Sciences de l'Univers of the Centre National de la Recherche Scientifique (INSU/CNRS) of France and the University of Hawaii, while TBL/NARVAL are operated by INSU/CNRS. We thank the CFHT and TBL staff for their help during the observations.


\appendix
\section{Runaway electron generation and the acceleration region}
\label{app:Dreicer}
The component of the electric field that is parallel to the magnetic field in the acceleration region is given by

\begin{equation}
	E=-\frac{2\sqrt{2\pi}}{\Gamma (1/4)}H^{-1/4}v_{eff}B_{\star}(r_{orb}),
\end{equation}
where $H$ is the magnetic Reynolds number and the remaining symbols are defined in the main text. This is the full version of Eq. (\ref{eq:EField}). \citet{Jardine2008} give a full derivation and discussion of this expression. In a plasma, the motions of the electrons, under the action of an electric field, are opposed by a drag force caused by coulomb collisions. For electrons with high enough velocities, the coulomb drag becomes negligible and the electrons can freely run out of the distribution. The exact behaviour is determined by the strength of the electric field. Below a critical field strength, known as the Dreicer field \citep{Dreicer1959}, only the electrons in the high velocity tail of the distribution runs away. However, at super-Dreicer strengths, the entire distribution is able to do so. The Dreicer field is given by

\begin{equation}
	E_D=18\times10^{-12}n_{cs}T^{-1},
	\label{eq:Ed}
\end{equation}
where the $n_{cs}$ and $T$ are the electron density and temperature respectively inside the current sheet. Due to compression of the plasma within the current sheet, the density does not necessarily take on their wind values. Both the compressed density and temperature in the current sheet are unknown. We find that an increase in the density by a factor of ~15 and a temperature of 1MK reproduces the emitted radio power at the solar system planets well and adopt these values. The rate at which runaway electrons is given by \citet{Kruskal1964}:

\begin{equation}
	\dot{n}_{run}=0.35n_{cs}\nu_cf(E/E_D) 
	=2.6\times10^{-5}n_{cs}^2T^{-3/2}f(E/E_D)
\end{equation}
where $\nu_c$ is the electron collision frequency and $f(E/E_D)$ is given by

\begin{equation}
	f\left(\frac{E}{E_D}\right)=\left(\frac{E_D}{E}\right)^{3/8}\exp\left[-\left(\frac{2E_D}{E}\right)^{1/2}-\frac{E_D}{4E}\right].
\end{equation}

\end{document}